\documentclass[12pt]{mn2e}
\usepackage{graphicx}
\usepackage{epsf}
\usepackage{lscape}
\usepackage{longtable}
\usepackage{rotating}
\usepackage{color}
\usepackage{comment}
\usepackage{subfigure}

\newcommand{\gtrsim}{\ga}
\newcommand{\lesssim}{\la}

\newcommand{\apjl}{ApJ}

\newcommand{\mnras}{MNRAS}

\begin{document}
\topmargin -0.5in 

\title[Spectroscopy of $z\sim7$ candidate galaxies]{Spectroscopy of $z\sim7$ candidate galaxies: Using Lyman-$\alpha$ to constrain the neutral fraction of hydrogen in the high-redshift universe.\thanks{Based on observations collected at the European Southern Observatory, Chile, as part of programmes 086.A-0968 and 088.A-1013.}}

\author[Joseph\ Caruana, et al.\ ]
{Joseph Caruana$^{1,2}$\thanks{E-mail: jcaruana@aip.de}, Andrew J. Bunker$^{2}$, Stephen M. Wilkins$^{2,3}$, Elizabeth R. Stanway$^{4}$, 
\newauthor
Silvio Lorenzoni$^{2,5}$, Matt J.\ Jarvis$^{2,6}$ \& Holly Ebert$^2$ \\
$^{1}$\,Leibniz-Institut f\"{u}r Astrophysik, An der Sternwarte 16, 14482 Potsdam \\
$^{2}$\,University of Oxford, Department of Physics, Denys Wilkinson Building, Keble Road, OX1 3RH, U.K. \\ 
$^{3}$\,Astronomy Centre, Department of Physics and Astronomy, University of Sussex, Brighton, BN1 9QH, U.K. \\
$^{4}$\,Department of Physics, University of Warwick, Gibbet Hill Road, Coventry, CV4 7AL, U.K. \\
$^{5}$\,Centro de Astronomia e Astrof\'{\i}sica da Universidade de Lisboa,  Observat\'{o}rio Astron\'{o}mico de Lisboa, Lisbon, Portugal \\
$^{6}$\,Physics Department, University of the Western Cape, Bellville 7535, South Africa \\
}

\maketitle

\begin{abstract}
Following our previous spectroscopic observations of $z>7$ galaxies with Gemini/GNIRS and VLT/XSHOOTER, which targeted a total of 8 objects, we present here our results from a deeper and larger VLT/FORS2 spectroscopic sample of Wide Field Camera 3 selected $z>7$ candidate galaxies.  With our FORS2 setup we cover the 737-1070nm wavelength range, enabling a search for Lyman-$\alpha$ in the redshift range spanning 5.06 - 7.80.  We target 22 $z$-band dropouts and find no evidence of Lyman-$\alpha$ emission, with the exception of a tentative detection ($<5\sigma$, which is our adopted criterion for a secure detection) for one object.  The upper limits on Lyman-$\alpha$ flux and the broad-band magnitudes are used to constrain the rest-frame Equivalent Widths for this line emission.  We analyse our FORS2 observations in combination with our previous GNIRS and XSHOOTER observations, and suggest that a simple model where the fraction of high rest-frame Equivalent Width emitters follows the trend seen at $z=3-6.5$ is inconsistent with our non-detections at $z\sim7.8$ at the 96\% confidence level.  This may indicate that a significant neutral HI fraction in the intergalactic medium suppresses Lyman-$\alpha$, with an estimated neutral fraction $\chi_{HI}\sim0.5$, in agreement with other estimates.
\end{abstract} 


\section{Introduction}

Recent years have seen many attempts at spectroscopically confirming candidate galaxies at $z > 7$ (e.g. Fontana et al.\ 2010, Pentericci et al.\ 2011, Vanzella et al.\ 2011, Caruana et al.\ 2012, Schenker et al.\ 2012, Ono et al.\ 2012).  There are three main desired objectives for such an endeavour.  The first goal is to investigate whether Lyman-$\alpha$ is able to emerge at all at these high redshifts, given that studies of the Gunn-Peterson optical depth in spectra of quasars at $z>6$ show that the fraction of neutral hydrogen ($\chi_{HI}$) is high enough ($>1$\%) to result in a complete Gunn-Peterson trough (Gunn \& Peterson, 1965) due to total absorption of Lyman-$\alpha$ photons (Becker et al. 2001, Fan et al. 2002, Fan et al. 2006), indicating that there is a large neutral fraction at higher redshifts.  Secondly, any detectable Lyman-$\alpha$ emission can be used to confirm the redshift of these objects, in so doing confirming the validity of the Lyman-break technique to reliably select objects at ever higher redshifts.  Thirdly, since the Intergalactic Medium (IGM) in the early universe is expected to be much more neutral, the resonant property of the line makes Lyman-$\alpha$ emitters a very useful tool to probe the era of reionization.  In particular, the observed fraction of Lyman-$\alpha$ emitters can be used to constrain $\chi_{HI}$ at high redshift.

Detecting Lyman-$\alpha$ emission at $z > 7$ with ground-based facilities has proven to be quite a difficult challenge, both because of technical hurdles (e.g. small near-infrared arrays, with low quantum efficiency and without multiplexed spectroscopy) and especially because of the faintness of any potential Lyman-$\alpha$ emission when compared to the bright and rapidly varying sky emission in the Near-Infrared region of the electromagnetic spectrum.  For this reason, it is imperative to obtain independent spectroscopy of candidate high-$z$ sources, especially in view of the fact that some objects claimed to exhibit Lyman-$\alpha$ emission (e.g. Pell\'{o} et al.\ 2004, Lehnert et al.\ 2010) could not be confirmed in subsequent observations (e.g. Weatherley, Warren \& Babbedge \ 2004, Bunker et al.\ 2012).

In the present study, we significantly increase the sample size considered in our previous study (Caruana et al.\ 2012), targetting a number of $z$-band dropouts and a few $i$-band dropouts (observed as filler targets) to assess the emergence of Lyman-$\alpha$ emission at high redshift.  

The structure of this paper is as follows.  We describe our observations and data reduction in Section \ref{sec:obs} and present the results of our spectroscopy in Section \ref{sec:results}.  Analysis of the results and further discussion, particularly regarding possible implications for constraints on the Neutral Fraction of Hydrogen at $z\sim 7$ are found in Section \ref{sec:analysis}.  Section 5 provides a comparison with other studies in the literature.  We present conclusions in Section \ref{sec:conclusions}.  We adopt a $\Lambda$CDM cosmology throughout, with $\Omega_{M}= 0.3$, $\Omega_{\Delta}=0.7$ and $H_{0} = 70$\,km s$^{-1}$Mpc$^{-1}$. All magnitudes are in the AB system (Oke \& Gunn 1983).

\section{OBSERVATIONS AND DATA REDUCTION}
\label{sec:obs}

\subsection{Observations with FORS2}

FORS2 is a multi mode (imaging, polarimetry, long slit and multi-object spectroscopy) optical instrument which works in the 330-1100 nm wavelength range and has a maximum field size of $6.8'\times6.8'$ covered by a mosaic of two 2k$\times$4k MIT CCDs (hereafter referred to as Chip 1 and Chip 2).  We note that the sensitivity of the instrument drops sharply beyond $1.0\,\mu$m; see Figs. \ref{fig:fors2fl_aa} and \ref{fig:fors2fl_ab}. It was used in Mask Exchange Unit (MXU) mode to observe a number of $z$-band and $i$-band dropouts.  For these observations, we used the 600z holographic grism and OG590 order separation filter, which allowed us to span the 737-1070nm region.  (The precise wavelength range covered for any given slit varies slightly depending on the position of the slit on the detector.)

The main targets for our FORS2 observations (ESO program numbers 086.A-0968 and 088.A-1013, PI: A. Bunker) were a number of $z$-band dropouts (Table \ref{fors2_zdrops_table}) which were identified via imaging with the Hubble WFC3 camera in searches conducted by Wilkins et al.\ (2011), Bouwens et al.\ (2011), and McLure et al.\ (2010). The selection was geometric, satisfying the constraint that the slits must be $6''$ long and cannot overlap, with higher priority\footnote{Only targets which had non-detection in the optical bands were selected, and those targets with a more reliable dropout colour were given priority.  In the event that two candidates ended up being assigned the same priority, the brighter target was chosen.} targets preferred, given a choice.  We also observed a number of $i$-band dropouts (Table \ref{fors2_idrops_table}), 3 of which were identified by Bunker et al.\ (2004), 2 by Bouwens et al.\ (2006) in the Hubble Ultra Deep Field (HUDF) and 11 were chosen from the Chandra Great Observatories Origins Deep Survey (GOODS)-South field, but in the case of $i$-drops the observations are very incomplete as these were intended to be ``filler objects''.  Where no high redshift source was in slit contention, several lower redshift sources, expected to be strong line emitters, were also placed on the mask as in-situ tests of slit alignment and line detection sensitivity.  These are not discussed further here.

The observations were carried out in service mode over two semesters.  The first set of observations (as part of programme 086.A-0968A) was made on 2010 October 27, 28, 29 \& 30.  These observations comprise 11 pairs of frames, with the telescope pointing nodded along the slit between each 1400 second frame in order to facilitate subtraction of night sky emission lines.  The second set of observations (as part of programme 088.A-1013A), carried out using an identically cut mask, was made on 2011 November 08 \& 16, 2011 December 03, 06, 10, 11, 14 \& 16 and on January 2012, and 2012 March 14, 26 \& 27. These observations comprise 22 pairs of nodded frames, with each frame again having an integration time of 1400 seconds.  The total integration time on the science targets was 25.67 hours.

An acquisition frame was first obtained to check the pointing.  The observations were dithered in an ABBA sequence at positions $+3''$ and $-3''$ from the central coordinates along the slit long axis (i.e. a `chop' size of $6''$) to enable background subtraction.  The average seeing during the observations was of $0.7''$.  Most observations were taken at low airmass, which varied between 1.001 and 1.297.  The vast majority of the frames had an airmass of about 1.0, so the effect of differential atmospheric dispersion is negligible.  (Only two frames were acquired at a higher airmass beyond the range quoted above; for these two frames, the airmass at the start and end of the integration varied between 1.609 - 1.818 and 1.868 - 2.176 respectively.)  The resolving power attained for these observations at a wavelength of 1$\mu$m was measured from sky emission lines to be $R=1800=\lambda/\Delta\lambda$.

\begin{table*}
\centering
\begin{tabular}{ | c | c | c | c | c | c | c | c | c |}
\hline
 Object & R.A. (J2000) & Dec (J2000) & $J_{AB}$ & $H_{AB}$ & $M_{1600}$ & Wavelength Range & Chip & Selection \\
& & & (F125W) & (F160W) & & Spanned / \AA & & Catalogs \\
\hline
\hline
 ERS.z.87209 & 03:32:29.53 & -27:42:04.5 & 26.84 & 27.04 & -19.90 & 7373 - 10169 & Chip 1 & 1, 2\\
\hline
 ERS.z.90192 & 03:32:24.08 & -27:42:13.9 & 26.48 & 26.73 & -20.21 & 7373 - 9720 & Chip 1 & 1, 2\\
\hline
 ERS.z.26813 & 03:32:22.93 & -27:44:09.9 & 26.83 & 26.71 & -20.23 & 7373 - 9679 & Chip 1 & 1, 2\\
 \hline
 ERS.z.87326 & 03:32:23.15 & -27:42:04.6 & 26.96 & 27.76 & -19.18 & 7373 - 9628 & Chip 1 & 1 \\
 \hline
 ERS.z.46030 & 03:32:22.66 & -27:43:00.7 & 26.11 & 25.95 & -20.99 & 7373 - 9620 & Chip 1 & 1, 2\\
 \hline
 ERS.z.70546 & 03:32:27.90 & -27:41:04.2 & 25.97 & 26.51 & -20.43 & 7373 - 10000 & Chip 1 & 1, 2 \\
 \hline
 HUDF.z.2677 & 03:32:42.18 & -27:46:27.9 & 27.70 & 27.72 & -19.22 & 8104 - 10995$^\dagger$ & Chip 2 & 1, 2, 3 \\
 \hline
 HUDF.z.4444 & 03:32:42.54 & -27:46:56.6 & 26.44 & 26.41 & -20.53 & 8155 - 10995$^\dagger$ & Chip 2 & 1, 2, 3, 4 \\
 \hline
 HUDF.z.6433 & 03:32:42.55 & -27:47:31.5 & 27.10 & 27.30 & -19.64 & 8181 - 10995$^\dagger$ & Chip 2 & 1, 2, 3, 4 \\
 \hline
 HUDF.z.7462 & 03:32:36.76 & -27:47:53.6 & 27.75 & 27.82 & -19.12 & 7720 - 10981$^\dagger$ & Chip 2 & 1, 2, 3 \\
 \hline
 HUDF.z.5141 & 03:32:38.79 & -27:47:07.2 & 26.90 & 26.76 & -20.18 & 7848 - 10995$^\dagger$ & Chip 2 & 1, 2, 3, 4  \\
 \hline
 HUDF.z.1889 & 03:32:41.81 & -27:46:11.3 & 28.44 & 28.91 & -18.03 & 8062 - 10987$^\dagger$ & Chip 2 & 1, 2 \\
 \hline
 HUDF.z.6497 & 03:32:36.45 & -27:47:32.4 & 28.31 & 28.18 & -18.76 & 7680 - 10938 & Chip 2 & 1, 2, 3 \\
 \hline
zD4 & 03:32:39.52 & -27:47:17.4 & 26.60 & 26.40 & -20.54 & 7916 - 10995$^\dagger$ & Chip 2 & 2, 3, 4 \\
 \hline
zD7 & 03:32:44.69 & -27:46:44.3 & 27.0 & 27.10 & -19.84 & 8340 - 10995$^\dagger$ & Chip 2 & 2, 3, 4 \\
 \hline
zD9 & 03:32:37.21 & -27:48:06.1 & 27.60 & 27.60 & -19.34 & 7800 - 10995$^\dagger$ & Chip 2 & 2, 3, 4 \\
 \hline
 UDFz-41597044 & 03:32:41.59 & -27:47:04.4 & 28.30 & 28.6 & -18.34 & 8080 - 10990 & Chip 2 & 2 \\
 \hline
 M2560z & 03:32:37.79 & -27:47:40.4 & 28.87 & 29.13 & -17.81 & 7795 - 10986$^\dagger$ & Chip 2 & 3\\
 \hline
 ERSz-2432842478 & 03:32:43.27 & -27:42:47.8 & 25.90 & 25.8 & -21.14 & 8083 - 10995$^\dagger$ & Chip 1 & 2 \\
 \hline
 ERSz-2354442550 & 03:32:35.43 & -27:42:55.0 & 26.10 & 26.20 & -20.74 & 7422 - 10692$^\dagger$ & Chip 1 & 2 \\
 \hline
 ERSz-2225141173 & 03:32:22.50 & -27:41:17.3 & 27.20 & 27.10 & -19.84 & 7373 - 9555 & Chip 1 & 2 \\
 \hline
 ERSz-2352941047 & 03:32:35.28 & -27:41:04.7 & 27.60 & 27.40 & -19.54 & 7373 - 10000 & Chip 1 & 2 \\
\hline
\end{tabular} 

$^\dagger$ For these galaxies, the spectral range falling on the CCD included the entire red end of the grism transmission. We note
the throughput of FORS2 drops sharply past 10000\,\AA .\\
\caption{$z$-drops targetted with FORS2.  The number in the last column denotes the catalogs in which these objects have been selected, where 1=Wilkins et al. (2011), 2=Bouwens et al.\ (2011), 3=McLure et al.\ (2010) and 4=Bunker et al.\ (2010).  The absolute magnitude at 1600\AA\, is computed assuming that the object lies at the peak of the expected redshift distribution for the $z$-drop selection (see Fig. \ref{fig:sims_fors2}).  }
\label{fors2_zdrops_table} 
\end{table*}

\begin{table*}
\centering
\begin{tabular}{ | c | c |}
\hline
Observed Target & Observed Target \\
in Our Catalog & in Other Literature \\
\hline
\hline
ERS.z.87209 & G2\_6173 in Fontana et al. (2010) \\
& ERS 7376 in Schenker et al. (2012) \\
\hline
ERS.z.90192 & W\_6 in Fontana et al. (2010) \\
\hline
ERS.z.46030 & G2\_2370 in Fontana et al. (2010) \\
\hline
HUDF.z.4444 & G2\_1408 in Fontana et al. (2010) \\
\hline
HUDF.z.6433 & O\_5 in Fontana et al. (2010) \\
\hline
\end{tabular} 
\caption{$z$-drops targeted by our spectroscopy which were also observed by other groups.  No Lyman-$\alpha$ emission was detected in the literature for any of these objects, except for HUDF.z.4444, which was reported by Fontana et al. (2010) to exhibit a tentative emission line which, if real, would place it at $z=6.97$.  Our observations with FORS2, however, do not confirm this detection.}
\label{observed_others} 
\end{table*}

\begin{table*} 
\centering 
\begin{tabular}{| c || c | c | c | c |}
\hline
Object & $z_{AB}$ (F850lp) & R.A. (J2000) & Dec. (J2000) & Absolute Magnitude \\
\hline
\hline
46574	&	26.71	& 03:32:38.28	& -27:46:17.2 & -19.98 \\
\hline
49117D	&	27.74	& 03:32:38.96	& -27:46:00.5 & -18.95 \\
\hline
42806	&	28.21	& 03:32:36.49	& -27:46:41.4 & -18.48 \\
\hline
HUDF-39065387	&	26.92	& 03:32:39.06	& -27:45:38.7 & -19.8383 \\
\hline
HUDF-35237429	&	29.23	& 03:32:35.23	& -27:47:42.9 & -17.53 \\
\hline
CDFS-2379542076	&	25.80	& 03:32:37.95	& -27:42:07.6 & -20.95 \\
\hline
CDFS-2452643595	&	26.08	& 03:32:45.26	& -27:43:59.5 & -20.68 \\
\hline
CDFS-2278843156	&	26.19	& 03:32:27.88	& -27:43:15.6 & -20.57 \\
\hline
CDFS-2294145379	&	26.35	& 03:32:29.41	& -27:45:37.9 & -20.41 \\
\hline
CDFS-2340645186	&	26.87	& 03:32:34.06	& -27:45:18.6 & -19.88 \\
\hline
CDFS-2323746215	&	26.92	& 03:32:32.37	& -27:46:21.5 & -19.84 \\
\hline
CDFS-2431845175	&	26.92	& 03:32:43.18	& -27:45:17.5 & -19.84 \\
\hline
CDFS-2373844457	&	26.98	& 03:32:37.38	& -27:44:45.7 & -19.78 \\
\hline
CDFS-2455245382	&	27.13	& 03:32:45.52	& -27:45:38.2 & -19.63 \\
\hline
CDFS-2418044023	&	27.23	& 03:32:41.80	& -27:44:02.3 & -19.53 \\
\hline
CDFS-2279641190	&	27.41	& 03:32:27.96	& -27:41:19.0 & -19.35 \\
\hline
\end{tabular} 
\caption{i-drops targeted with FORS2.  The first three objects are from Bunker et al.\ (2004) and the rest are from Bouwens et al.\ (2006).}
\label{fors2_idrops_table} 
\end{table*}

\subsection{Data Reduction}

The first stages of the data reduction were carried out using the ESO FORS2 pipeline\footnote{The FORS2 pipeline is available here:  http://www.eso.org/sci/software/pipelines/fors/fors-pipe-recipes.html}.  In particular, the two pipeline recipes {\tt fors\_calib} and {\tt fors\_science} were used for the preparation of the calibration frames and science data respectively.  In this section, we first review what the pipeline-reduction entailed, then move on to discuss the later stages of the reduction process which required custom-made procedures.

The initial reduction steps involved using the pipeline recipe {\tt fors\_calib} for three main purposes: (a) identifying reference lines on our MXU arc lamp exposures to perform wavelength calibration, (b) tracing the spectral edges on the flat field exposures, and (c) creating a normalized flat-field from a number of input flat-field exposures.

The pipeline recipe {\tt fors\_science} was then used to separately reduce each raw frame containing the science spectra, applying the normalized master flat-field and extraction mask created by {\tt fors\_calib}.  It was also used to perform bias-subtraction and flat-fielding of the data.  The flat-fielding procedure entailed dividing the bias subtracted input scientific frames by the normalised flat field frame.  The raw science spectra were remapped to a spectral pixel scale of 1.63\AA/pix and a spatial pixel scale of $0.25''$/pix; the remapping carried out by this recipe eliminates optical distortions, which manifest themselves in problems such as spatial curvature in the raw frames.  Local sky subtraction was performed by modelling the sky for each column of pixels for each spectrum.  This was done prior to resampling the data in order to minimize small-scale interpolation issues.  Cosmic rays were also eliminated using this recipe.

The combination of individual science frames was not carried out using the FORS2 pipeline.  Instead, we wrote our own algorithm to carry out this task.  In particular, since the exact start and end y-coordinates of a given reduced spectrum in the pipeline output sometimes varied by about a pixel between frames (the pipeline introduced a padding between individual slits which was not always constant), this tailor-made algorithm was used to read in and calculate the relevant information about each reduced spectrum (e.g. the length of the slit for that spectrum, the central y-coordinate of the slit, etc.) to cut each spectrum correctly and enable the combination of frames.  The actual combination was carried out using the {\tt IRAF} task {\tt imcombine}, averaging over the number of frames.

One of the slits of our mask was placed on an M4III-star (EIS J033236.27-274302.7) and observations of this star were used to flux-calibrate our data.  We derived the instrument efficiency and a flux calibration curve by comparing the observed spectrum of this star against a theoretical spectrum of an M4III star.  A smooth function was then fit through the calibration curve using {\tt IRAF}, taking into account strong atmospheric absorption by O$_{2}$ and H$_{2}$O around $0.76\mu$m.  Flux calibration was obtained via convolution of the spectral response curve with the F775W $i$-band filter throughput model and a comparison to the star's known apparent magnitude.

\section{Results}
\label{sec:results}

\subsection{No Lyman-$\alpha$ emission in $z$-band dropouts}

\begin{figure}
	\subfigure{
	\includegraphics[width=\linewidth]{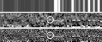}
    }
   \subfigure{ 
    \includegraphics[width=\linewidth]{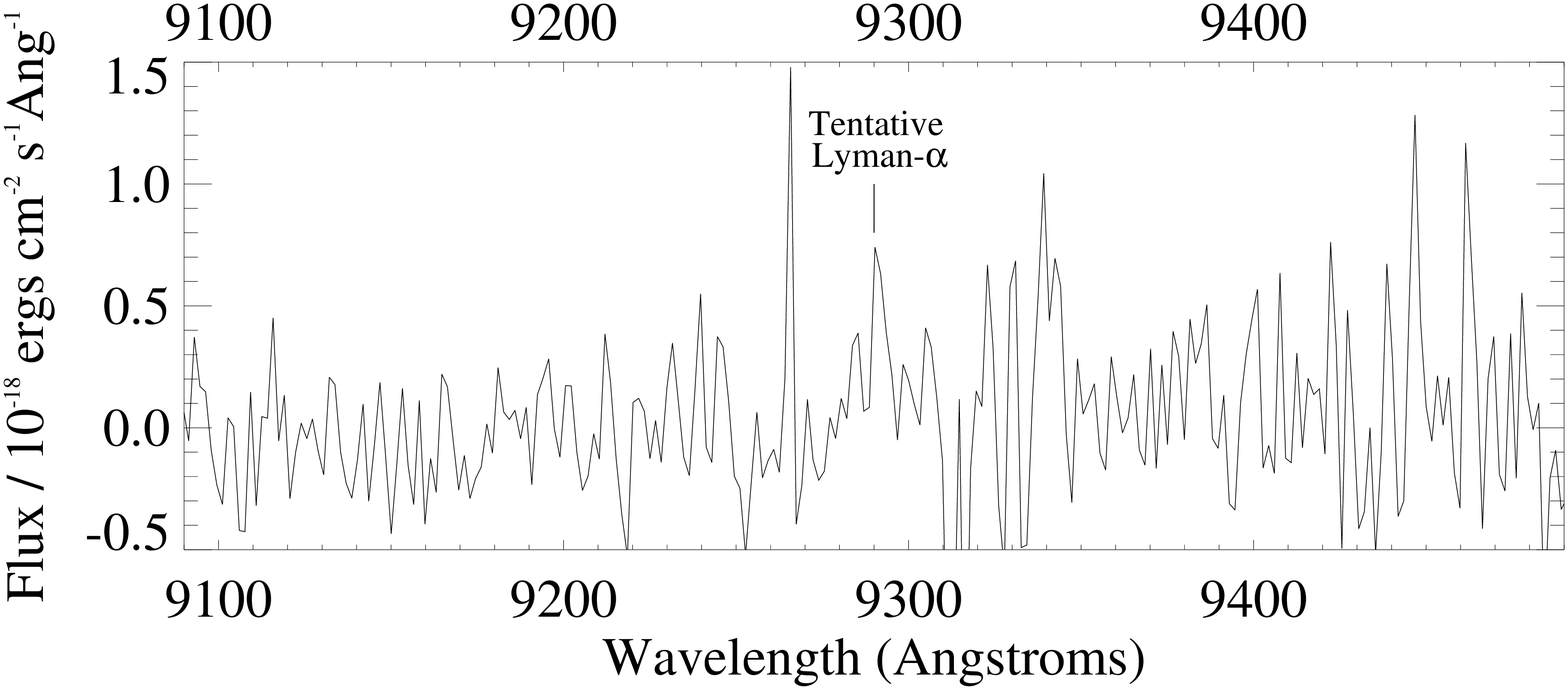}
    }
    \caption{ERSz-2225141173.  Top:  The lower panel shows 2D spectrum (normalised by the noise) around Lyman-$\alpha$ with the emission feature circled (the horizontal line below the emission line is not continuum but merely a slit-edge effect), the middle panel shows the same smoothed with a Gaussian of $\sigma=1$pix, and the upper panel shows the sky spectrum for the same wavelength range.  Bottom:  The 1D spectrum around Lyman-$\alpha$, extracted over a width of 1.25$''$ (5 pixels).  The observed Lyman-$\alpha$ emission line is centered on 9290\AA\, placing the object at a redshift $z=6.64$.  This tentative emission line is detected at $< 5\sigma$.}
    \label{fig:possible_detection}
\end{figure}

No Lyman-$\alpha$ emission was detected in any of the spectra for this sample of $z\sim7$ galaxies, with the exception of a tentative Lyman-$\alpha$ emission line for ERSz-2225141173 (see Fig. \ref{fig:possible_detection}) which, at 4.8$\sigma$, lies just below our nominal 5$\sigma$ Lyman-$\alpha$ detection threshold.  (This feature was also independently detected in our data by Pentericci et al.\ 2014.) Since the continuum flux for these objects is known from {\emph HST} imaging, this was used in conjunction with the spectroscopy to derive upper limits to the Equivalent Width of these objects (e.g. Figure \ref{fig:fors2ew_aa}) as described in detail in Section \ref{sec:analysis}.

We note that amongst our sample of targeted galaxies, 5 objects have also been observed by other groups, as listed in Table \ref{observed_others}.  None of these objects have any claimed Lyman-$\alpha$ detection in the literature except for HUDF.z.4444.  In Caruana et al.\ (2012), particular attention was paid to this object (referred to in that study as HUDF.zD1 from the catalog of Bunker et al.\ 2010), a target which had been previously observed by Fontana et al.\ (2010), their target G2\_1408, who note a marginal detection of possible Lyman-$\alpha$ emission at a wavelength of 9691.5\AA.  If this emission line were real it would place this object at $z=6.97$.  This object was observed with Gemini/GNIRS and no emission was detected.  However, as was noted in Caruana et al.\ (2012), our spectroscopy with GNIRS was not deep enough to comment with certainty about the reality of the line.  This target was observed again with FORS2 and we do not find concrete evidence of Lyman-$\alpha$ emission at the reported wavelength.  While we do obtain a measure of 3.22$\sigma$ in the location on the 2D-spectrum where the claimed Lyman-$\alpha$ emission would lie, this is insufficient to report a confirmed detection (as can also be seen in the first and third panels of Fig. \ref{fig:fontana_fors2}).  Furthermore, inserting an artificial Lyman-$\alpha$ line with the same strength reported in Fontana et al. (2010), we expect that this would be detected at a level of 10$\sigma$ (see last panel of Fig. \ref{fig:fontana_fors2}).  We also searched for Lyman-$\alpha$ emission in other wavelength regions, but find none.  Therefore, we rule out the earlier reported line flux at a level $>3 \sigma$ ($>99\%$ probability).

\begin{figure}
   \resizebox{0.50\textwidth}{!}{\includegraphics{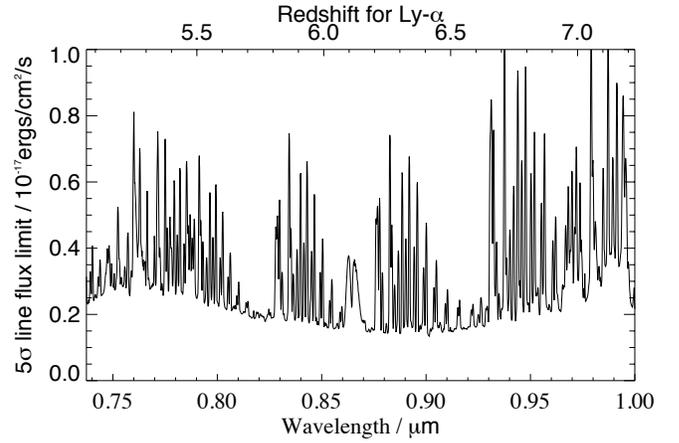}} \\
 \caption{A representative 5$\sigma$ line flux limit plot from our FORS2 observations, here showing the case of a slit placed on the chip such that the wavelength range spans the region between $\sim0.74\mu$m - $1.0\mu$m.  (The object considered here is ERS.z.70546.)}
 \label{fig:fors2fl_aa}
\end{figure}

\begin{figure}
   \resizebox{0.50\textwidth}{!}{\includegraphics{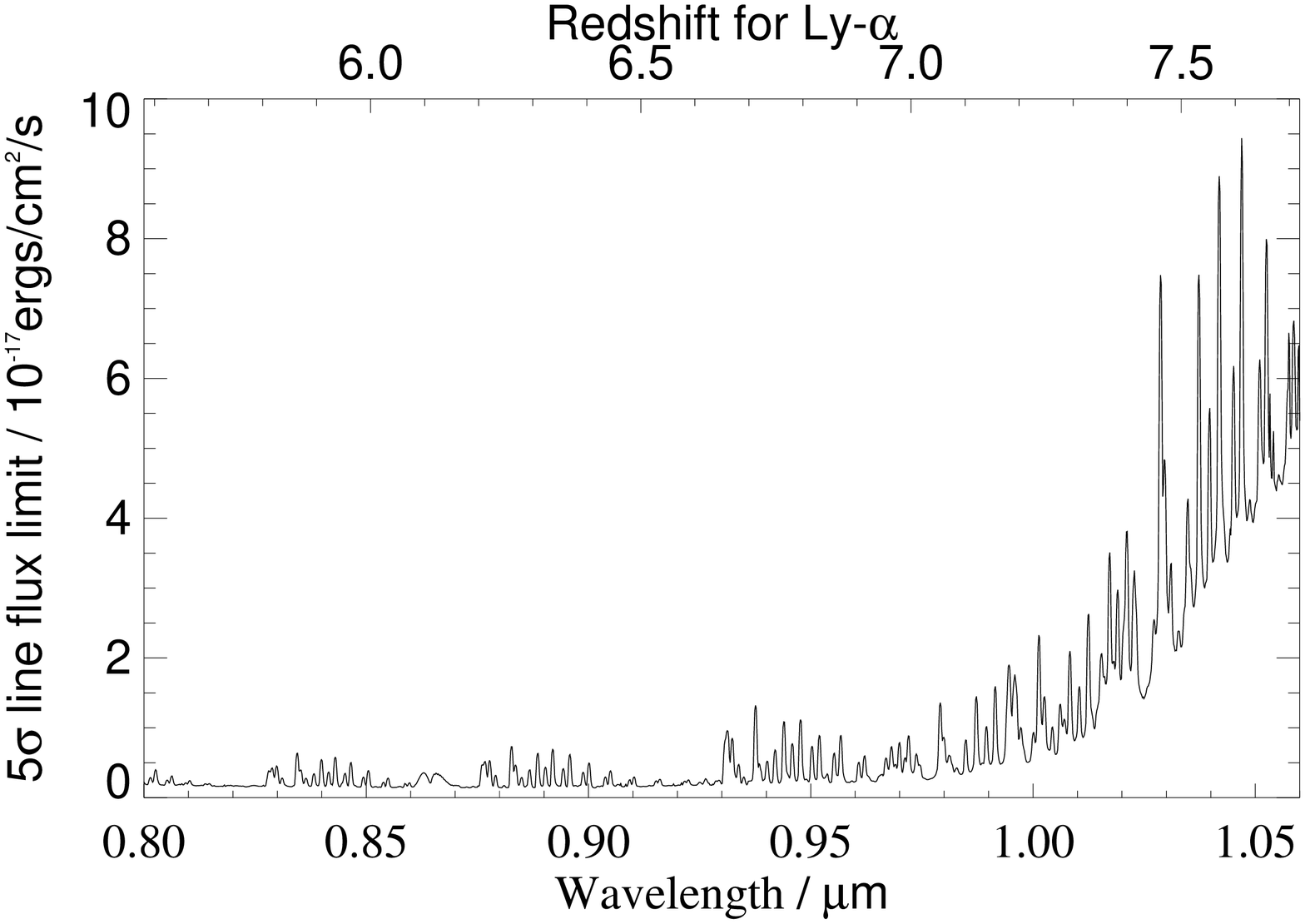}} \\
 \caption{A representative 5$\sigma$ line flux limit plot from our FORS2 observations, here showing the case of a slit placed on the chip such that the wavelength range spans the region between $\sim0.79\mu$m - $1.1\mu$m.  (The object considered here is UDFy-39537174.)}
 \label{fig:fors2fl_ab}
\end{figure}

\begin{figure}
   \resizebox{0.50\textwidth}{!}{\includegraphics{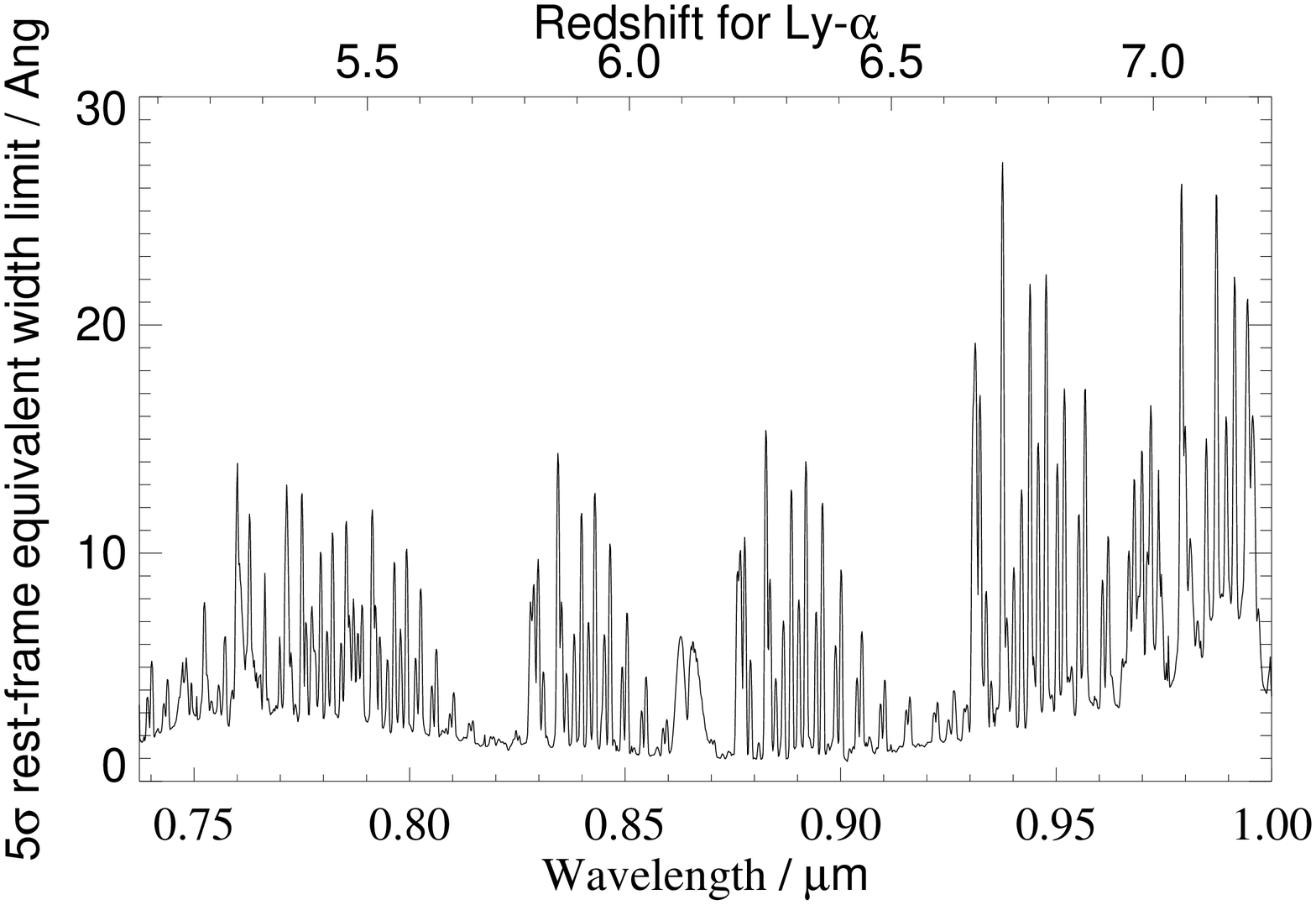}} \\
 \caption{A representative 5$\sigma$ Rest-Frame Equivalent Width limit plot from our FORS2 observations, here showing for the case of ERS.z.70546.}
 \label{fig:fors2ew_aa}
\end{figure}

\begin{figure}
   \resizebox{0.50\textwidth}{!}{\includegraphics{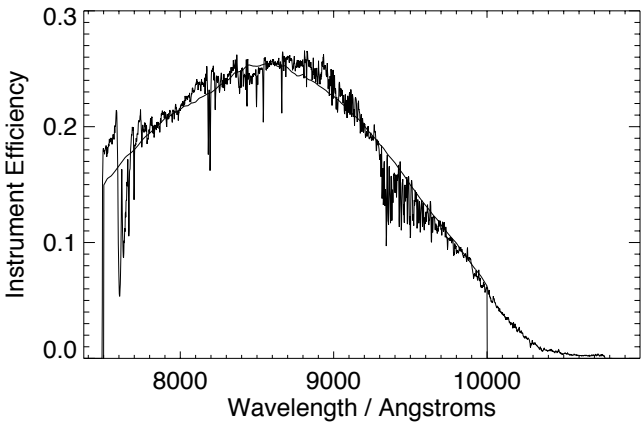}} \\
 \caption{Our derived efficiency curve for FORS2 compared with that quoted by ESO (smooth curve).}
  \label{fig:efficiency_fors2}
\end{figure}

\begin{figure}
\begin{center}
  \resizebox{0.47\textwidth}{!}{\includegraphics{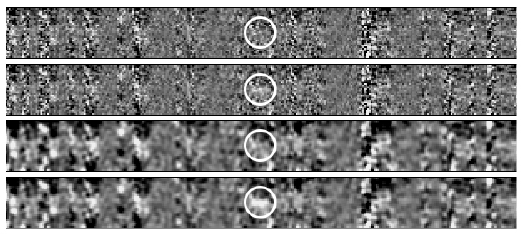}} \\
\end{center}
 \caption{The calibrated FORS2 spectrum, with the location
 of HUDF.z.4444 and the expected wavelength of the tentative Lyman-$\alpha$ emission reported 
 by Fontana et al.\ (2010) marked with a white circle. Wavelength increases from left to right.
 From top to bottom: (a) the reduced data. Vertical lines of higher noise are due to night sky emission lines; 
 (b) an artificial source with the same line flux ($3.4\times 10^{-18}\,{\rm erg\,cm^{-2}\,s^{-1}}$), wavelength and profile as the Fontana et al.\ (2010) line added into the frame;
 (c) the reduced data convolved with an elliptical Gaussian with $\sigma=1$;
 (d) an artificial source with the same line flux ($3.4\times 10^{-18}\,{\rm erg\,cm^{-2}\,s^{-1}}$)
 and wavelength as the Fontana et al.\ 2010 line added into the frame. The resulting frame has been Gaussian smoothed ($\sigma=1$).  A line with the same strength as that reported in Fontana et al.\ 2010 (which was measured to have S/N $\leq$\, 7 in their spectroscopy) is expected to be detected at 10$\sigma$ in our observations, but only 3.22$\sigma$ is measured.}
 \label{fig:fontana_fors2}
\end{figure}

\subsection{Detection of Lyman-$\alpha$ emission in 4 $i$-band dropouts}

\begin{figure}
	\subfigure{
	\includegraphics[width=\linewidth]{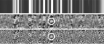}
    }
   \subfigure{ 
    \includegraphics[width=\linewidth]{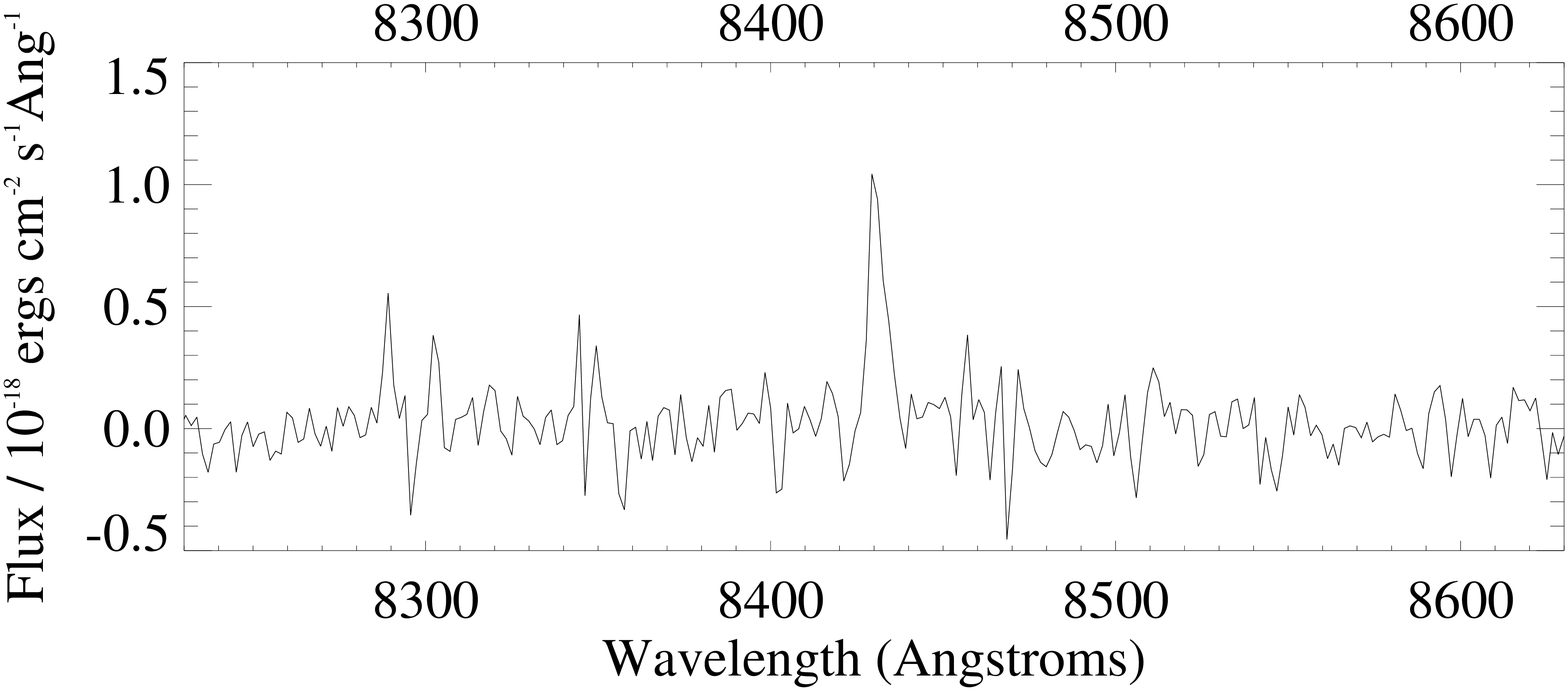}
    }
    \caption{CDFS-2418044023.  Left:  The lower panel shows 2D spectrum around Lyman-$\alpha$ (with the emission feature circled), the middle panel shows the same smoothed with a Gaussian of $\sigma=1$pix, and the upper panel shows the sky spectrum for the same wavelength range.  Right:  The 1D spectrum around Lyman-$\alpha$, extracted over a width of 1.25'' (5 pixels).  The observed Lyman-$\alpha$ emission line is centered on 8430.94\AA\, placing the object at a redshift $z=5.94$.  The total flux contained within the line is $5.13\times10^{-18}$ ergs cm$^{-2}$s$^{-1}$.  The Equivalent Width of the line is 38\AA.}
     \label{fig:our1}
\end{figure}

Our spectroscopy also targetted 16 $i$-band dropouts; these were observed as filler targets, and as such are not the main subject of this study.  Amongst these, Lyman-$\alpha$ emission is detected in 4 objects, enabling a precise calculation of the redshift for these galaxies.  The spectroscopic redshift of two of these objects had already been determined before by Stanway et al.\ (2004) and Vanzella et al.\ (2009), whereas the redshift for two other objects is being presented here for the first time.  The properties of these objects are summarised in Table \ref{idrop_table} and Figures \ref{fig:our1} - \ref{fig:vanzella_line}.  The main other plausible line would be [OII] 3727 at about $z=1.2$, but the $i-z$ colours of all the dropouts targetted are $(i-z)>1.5$ AB magnitudes (Bouwens et al.\ 2006, Bunker et al. 2004), meaning a drop in flux density of more than a factor of 4 across these adjacent filters (F775W $i$-band and F850LP $z$-band). This scale of break is much more readily accomplished with the Lyman alpha forest break than a 4000\AA\,/Balmer break and dust reddening, which makes the Lyman-$\alpha$ interpretation much more plausible. Further weight is added by the non-detection in the F606W ``V-band" (which would be the U-band for the [OII] $z\sim1.2$ interpretation, where detectable rest-UV flux would be expected from the star formation powering the [OII] line).  

We note that the characteristic asymmetry often exhibited by the Lyman-$\alpha$ line at high redshift is observed for CDFS-2418044023 and HUDF-39065387 (Figs. 3 and 6 respectively) but not for CDFS-2373844457 and CDFS-2431845175 (Figs. 4 and 5) which have fainter line-fluxes (and hence lower signal to noise) than the other two.  This line-asymmetry is caused by intervening neutral hydrogen absorbing radiation shortward of 1216\AA\,, but as shown in Dayal et al. (2008), it can be reduced by gas inflows.

We note that our typical $5\,\sigma$ line flux limit around 8500\,\AA\ (corresponding to Lyman-$\alpha$ at $z\approx 6$) is $2.5\times 10^{-18}\,{\rm erg\,cm^{-2}\,s^{-1}}$ (see Figure~\ref{fig:fors2fl_aa}),
which corresponds to a rest-frame equivalent width of $\approx 15$\,\AA\ for our average magnitude of $z_{AB}\approx 27.0$ for the $i$-drops.
We compare our  Lyman-$\alpha$ fraction of 25\% (4 out of 16) with other work:
Pentericci et al. (2011) find 11 Lyman-alpha emitters in a sample of 17 drawn from a sample with $z_{AB}\approx 26.5$, but only 4 of these have rest-frame equivalent widths above our typical sensitivity of 15\,\AA ,
so the observed fractions are consistent at $\approx 25$\%. Stark et al.\ (2010)  have a similar range in absolute magnitude to our targetted $i$-drops, and present a Lyman-alpha fraction of 0.25 (28 detections amongst a sample of 108), which is the same as our own. We note that Curtis-Lake et al.\ (2011) target much brighter $i$-drops (with $z_{AB}<26$) and have a higher fraction of Lyman-$\alpha$ emission, $40-50$\%, but they note
that their $L>2\,L^*$ sample displays twice the strong Lyman-$\alpha$ emission fraction of samples with $L<L^*$ also at $z\approx 6$, which is again consistent with our work on the lower-luminosity sources.

\begin{table*}
\centering
\begin{tabular}{ | c | c | c | c |}
\hline
Observed Target & Redshift & EW (\AA) & Flux\\
\hline
\hline
CDFS-2418044023 & 5.94 & 38 & 5.13\\
\hline
CDFS-2373844457 & 6.08 & 15 & 3.50\\
\hline
CDFS-2431845175 & 5.93 & 24 & 4.76\\
\hline
HUDF-39065387 & 5.92 & 49 & 8.22\\
\hline
\end{tabular} 
\caption{$i$-drops targeted by our spectroscopy.  Units of flux are in $10^{-18}$ ergs cm$^{-2}$ s$^{-1}$.}
\label{idrop_table} 
\end{table*}

\begin{figure}
   \subfigure{
    \includegraphics[width=\linewidth]{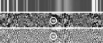}
    }
   \subfigure{ 
    \includegraphics[width=\linewidth]{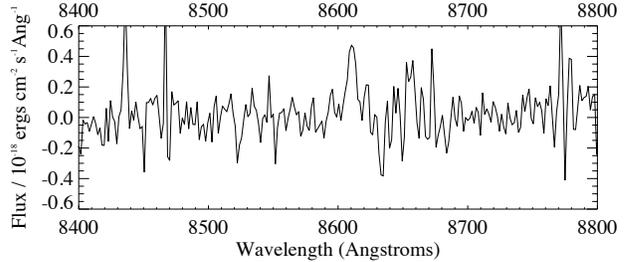}
    }
    \caption{The same as Figure \ref{fig:our1} for CDFS-2373844457.  The observed Lyman-$\alpha$ emission line is centered on 8610.58\AA\, placing the object at $z=6.08$.  The total flux contained within the line is $3.50\times10^{-18}$ ergs cm$^{-2}$s$^{-1}$.  The Equivalent Width of the line is 15\AA.}
\end{figure}

\begin{figure}
   \subfigure{
    \includegraphics[width=\linewidth]{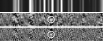}
    }
   \subfigure{ 
    \includegraphics[width=\linewidth]{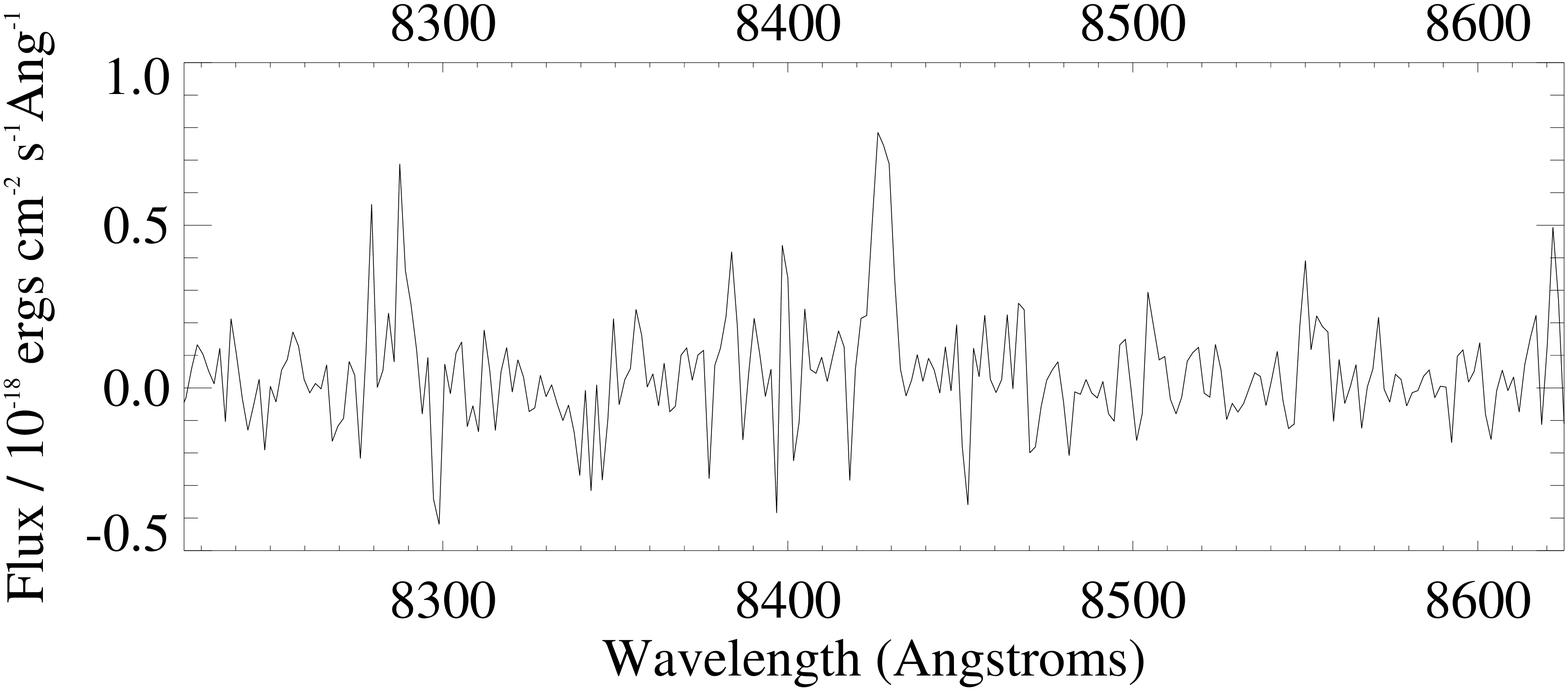}
    }
    \caption{The same as Figure \ref{fig:our1} for CDFS-2431845175.  The observed Lyman-$\alpha$ emission line is centered on 8427.05\AA,\ placing this object at $z=5.93$.  The total flux contained within the line is $4.76\times10^{-18}$ ergs cm$^{-2}$s$^{-1}$.  The Equivalent Width of the line is 24\AA.  This object has also been observed by Stanway et al.\ 2004 (Glare-3011).}
\end{figure}

\begin{figure}
   \subfigure{
    \includegraphics[width=\linewidth]{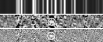}
    }
   \subfigure{ 
    \includegraphics[width=\linewidth]{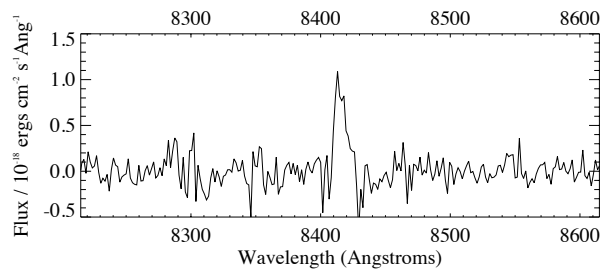}
    }
    \caption{The same as Figure \ref{fig:our1} for HUDF-39065387.  The observed Lyman-$\alpha$ emission line is centered on 8415.68\AA\, placing this object at $z=5.92$.  The total flux contained within the line is $8.22\times10^{-18}$ ergs cm$^{-2}$s$^{-1}$.  The Equivalent Width of the line is 49\AA.  This object has also been observed by Vanzella et al.\ 2009.}
      \label{fig:vanzella_line}
\end{figure}

\begin{table*}
\centering
\begin{tabular}{ | c | c | c | c | c | c | c | }
\hline
Object & Formal z-Range spanned & Frac$_{z}$ & Median EW & Frac$_{\mathrm{EW < 50\AA}}$ & Frac$_{\mathrm{EW < 75\AA}}$ & Frac$_{\mathrm{EW < 120\AA}}$ \\
& by data (for Ly-$\alpha$) & & & & & \\
\hline
\hline
ERS.z.87209 & 5.06 - 7.36 & 0.9516 & 11.85 & 0.9557 & 0.9916 & 1.000 \\
\hline
ERS.z.90192 & 5.06 - 7.00 & 0.5353 & 26.69 & 0.7111 & 0.8320 & 0.9283 \\
\hline
ERS.z.26813 & 5.06 - 6.96 & 0.5065 & 8.728 & 0.9940 & 1.000 & 1.000 \\
\hline 
*ERS.z.87326 & 5.06 - 6.92 & 0.5561& 8.634 & 0.9965 & 1.000 & 1.000 \\
\hline
ERS.z.46030 & 5.06 - 6.91 & 0.4077 & 2.959 & 1.000 & 1.000 & 1.000 \\
\hline
ERS.z.70546 & 5.06 - 7.23 & 0.7360 & 4.521 & 1.000 & 1.000 & 1.000 \\
\hline
HUDF.z.2677 & 5.67 - cutoff & 0.9853 & 94.11 & 0.4558 & 0.5570 & 0.6577 \\
\hline
HUDF.z.4444 & 5.71 - cutoff & 0.9998 & 29.15 & 0.6919 & 0.7443 & 0.8003 \\
\hline
HUDF.z.6433 & 5.73 - cutoff & 0.9940 & 58.25 & 0.5582 & 0.6579 & 0.7356 \\
\hline
HUDF.z.7462 & 5.35 - cutoff & 0.9816 & 72.09 & 0.4654 & 0.5539 & 0.6515 \\ 
\hline
HUDF.z.5141 & 5.46 - cutoff & 0.9992 & 58.45 & 0.5733 & 0.6708 & 0.7510 \\
\hline
HUDF.z.1889 & 5.63 - cutoff & 0.9892 & 137.1 & 0.3522 & 0.5754 & 0.7367 \\
\hline
HUDF.z.6497 & 5.32 - cutoff & 0.9841 & 139.0 & 0.2007 & 0.3692 & 0.5247 \\
\hline
zD4 & 5.51 - cutoff & 0.9998 & 37.08 & 0.6291 & 0.6920 & 0.7742 \\
\hline
zD7 & 5.86 - cutoff & 0.9966 & 381.3 & 0.0000 & 0.0947 & 0.2561 \\
\hline
zD9 & 5.42 - cutoff & 0.9916 & 87.11 & 0.4134 & 0.5451 & 0.6641 \\
\hline
*UDFz-41597044 & 5.65 - cutoff & 0.9941 & 119.5 & 0.3721 & 0.5513 & 0.7077 \\
\hline
*M2560z & 5.41 - cutoff & 0.9941& 209.1 & 0.0177 & 0.3242 & 0.5937 \\
\hline
*ERSz-2432842478 & 5.65 - cutoff & 0.9928 & 12.73 & 0.8687 & 0.9250 & 0.9666 \\
\hline
*ERSz-2354442550 & 5.11 - cutoff & 0.9877 & 13.77 & 0.8424 & 0.9027 & 0.9635 \\
\hline
*ERSz-2225141173 & 5.06 - 6.86 & 0.6886 & 11.59 & 0.9902 & 1.000 & 1.000 \\
\hline
*ERSz-2352941047 & 5.06 - 7.23 & 0.8299 & 28.48 & 0.7470 & 0.8884 & 0.9765 \\
\hline
\end{tabular} 
\caption{This table for all objects targeted by our spectroscopy shows the formal redshift range spanned by our data for Lyman-$\alpha$ in Column 2.  (Note that this column gives the formal spectral range, and where denoted by `cutoff' the spectra nominally extend to 1.1\,$\mu$m on the CCD, although we note that the sensitivity of FORS2 drops sharply beyond $\sim1\mu$m.)  Column 3 shows the fractional probability that a galaxy drawn from the dropout sample would fall within the formal spectral range of that particular spectrograph setup.  Column 4 gives the median EW for each object for the most probable redshift range (see Figure \ref{fig:sims_fors2}).  The remaining three columns tabulate the fraction of our spectroscopy that has EW limits lower than our chosen threshold (50\,\AA, 75\,\AA\, and 120\,\AA\, respectively.)  The figures in this table are for a line with an intrinsic velocity width of 300\,km\,s$^{-1}$.  An asterisk (*) before the object name indicates a marginal candidate which was not included in the chosen sample used for our analysis.}
\label{fors2_fraction_table} 
\end{table*}

\begin{table*}
\centering
\begin{tabular}{| c  c  c |}
\hline
& & \\
& $N_{\mathrm{eff}}=\sum$ Frac$_{z} \times$ Frac$_{\mathrm{EW < thres}}$  & \\
& & \\
\hline
EW$_{\mathrm{thres}}=50$\,\AA & EW$_{\mathrm{thres}}=75$\,\AA & EW$_{\mathrm{thres}}=120$\,\AA \\
\hline
7.247 &  8.459 & 9.601 \\
\hline
\end{tabular}
\caption{This table shows the total effective number of sampled galaxies with an EW upper limit lower than a set threshold, considering those $z$-drops targetted by FORS2 which were strong candidates.  The figures in this table are for a line with an intrinsic velocity width of 300\,km\,s$^{-1}$.}
\label{table2_fors2}
\end{table*}

\section{Analysis:  The EW distribution and The Lyman-$\alpha$ Fraction}
\label{sec:analysis}

All the 2D spectra were carefully inspected; in particular, we rigorously examined the expected spatial location for the target and the wavelength range around $0.9 - 1$ micron, where Lyman-$\alpha$ might be expected for $z$-band dropouts. We did this by means of visual inspection, including examining frames which had been convolved with a Gaussian with similar FWHM to the spatial seeing and spectral resolution to bring up any faint feature.  We then developed a noise model, based on the Poisson counts of the sky background and dark current, the sensitivity of the detector (as a function of wavelength and position on the array) and the readout noise of the array.   As a further check on the detectability of potential Lyman-$\alpha$ emission, artificial sources of varying intensity profiles were added in random locations in the 2D spectra, and we concluded that a line with an intrinsic velocity width of 300km\,s$^{-1}$ with a signal to noise ratio of 5$\sigma$ would be unambiguously detected in our spectroscopy using an aperture of size 1$''\times$18\AA\, (4$\times$11 pixels).

The continuum flux derived for the targetted objects from {\emph HST} imaging was used in conjunction with our deep spectroscopy to derive upper limits on the rest-frame Equivalent Widths (see Figure \ref{fig:fors2ew_aa} which shows the upper limit on the EW for a ] object in our sample).  The continuum flux was inferred from broad-band photometry using the filter above the Lyman-$\alpha$ break, using the F105W $Y$-band for the $z$-drops at $z\approx7$.  Above the Lyman break, the rest-frame spectral slope was modelled using $f_{\lambda}\propto\lambda^{\beta}$, where $\beta$ was taken to be -2 (Stanway, McMahon \& Bunker 2005; Wilkins et al.\ 2011).  Since at the high redshift extreme of the expected redshift distribution, the Lyman-$\alpha$ break can fall within the short-wavelength side of the filter used to determine the continuum, the upper limits on the line flux were used to correct for possible Lyman-$\alpha$ line emission contamination of the broad-band magnitude.

Since Lyman break galaxies are typically unresolved in ground-based observations, a spatial extraction aperture of 1.5 times the seeing disk was adopted.  For the spectral extent, a hypothetical line was assumed to have an intrinsic velocity width of 300km\,s$^{-1}$, which after convolution with the spectral resolution of FORS2 (5.48\AA) results in a line width of 11.4\AA\,FWHM ($\sim 7$ pixels).  Therefore, we adopted an aperture with size $1''\times18$\AA\,FWHM ($4\times11$ pixels).  For a 300km\,s$^{-1}$ line our aperture captures 93\% of the flux.

In the absence of detectable Lyman-$\alpha$ emission in the targetted $z$-drop sample, their precise redshifts cannot be determined. The $z\sim7$ identification for these sources is thus dependent on Lyman break colours in \emph{HST} photometry (as reported elsewhere). In further analysis, we thus assume the redshift distribution determined from simulations of Wilkins et al.\ (2011). For each spectroscopic target we computed M$_{1600}$, the absolute magnitude at $\lambda_{rest}=1600$\AA. The probability of recovering a galaxy of this magnitude, using our selection criteria, was computed as a function of redshift based on the Wilkins et al.\ (2011) simulations.

We then considered three different thresholds on the rest-frame EW for each target, namely 50\AA, 75\AA\, and 120\AA, and calculated the fraction of our spectroscopy which probed EW limits lower than a given threshold, (i.e. the fraction of the spectrum where EW$_{\mathrm{upper limit}} < $ EW$_{\mathrm{threshold}}$, which we denote by Frac$_{\mathrm{EW < thres}}$), and weighted this by the redshift probability distribution for the dropout galaxy.

\begin{figure}
\begin{center}
   \resizebox{0.47\textwidth}{!}{\includegraphics{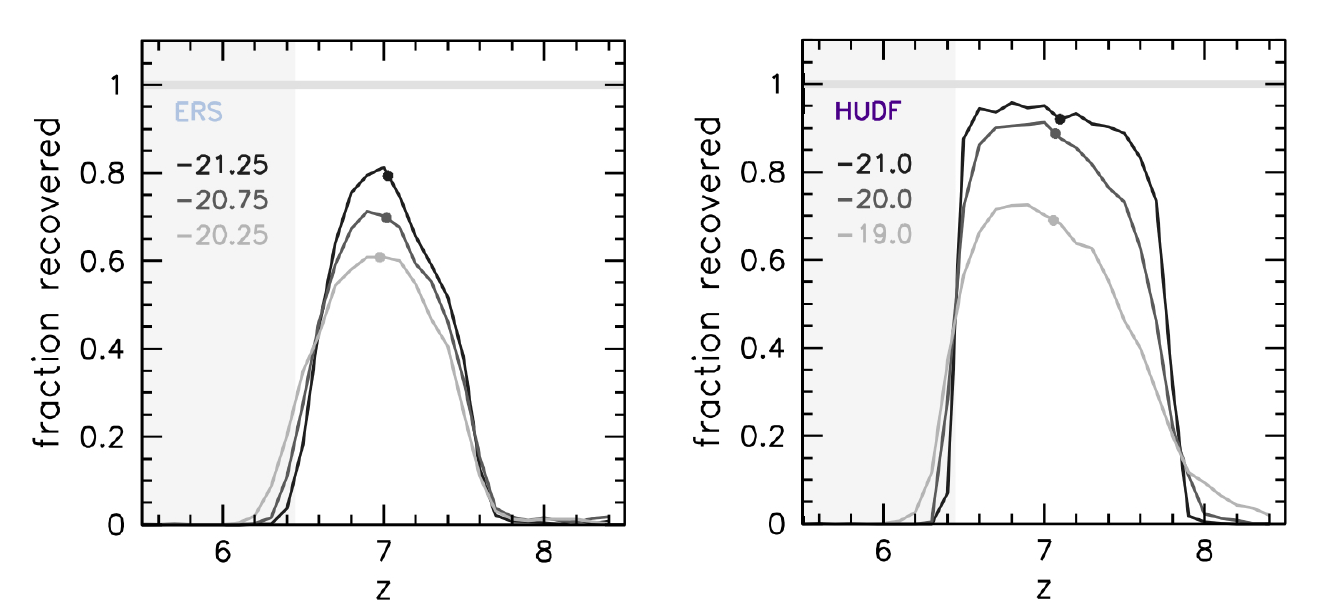}} \\
\end{center}
 \caption{The probability of recovering a galaxy in the simulations for $z$-band dropouts described in Wilkins et al.\ (2011) for the HUDF and ERS fields as a function of redshift for several different absolute rest-frame M$_{1600}$ magnitudes.  The results of these simulations are used to provide the expected redshift distributions for our $z$-drops.  In both figures, the mean redshift is denoted by a dot.} 
 \label{fig:sims_fors2}
\end{figure}

The position of each slit on the detector varied; this means that the spectral range covered is different for each slit.  The likelihood function for a galaxy to be lying at a certain redshift enabled us to compute the fractional probability (Frac$_z$) that a galaxy drawn from the sample of observed targets would fall within the spectral range covered.  The product of Frac$_z$ and Frac$_{\mathrm{EW<thresh}}$ gives the effective number of observed galaxies, $N_{\rm eff}$, probed with a sensitivity greater than the chosen EW threshold.  Table \ref{table2_fors2} gives $N_{\rm eff}$ for our three chosen thresholds. To derive these numbers we did not consider all our spectroscopically targetted objects.  Instead, we chose only those robust targets which appear in more than one catalog of Lyman-break galaxies selected by different groups.  The less secure objects which were excluded from our analysis are marked with an asterisk (*) in Table \ref{fors2_fraction_table}.  Also, it should be noted that the photometric redshift selection of McLure et al.\ (2010) reduces to a standard Lyman break selection for the most secure objects.  So our main selection requirement was a significant spectral break which is, with high probability, the Lyman break at high redshift, and colours in the other available wavebands which ruled out with high probability a lower-redshift ``red galaxy" or Galactic star interpretation.

Given that no Lyman-$\alpha$ emission is detected in the data, what scenarios of EW evolution from lower redshift can be ruled out?  As described in our previous paper, fom Poisson statistics, if a given model predicts that on average $\lambda_{\rm ex}$ galaxies are expected to be detected in the survey, the probability $f_{n}$ of detecting $n$ galaxies is given by:
\begin{equation}
f_{n}=\frac{(\lambda_{\rm ex})^n e^{-\lambda_{\rm ex}}}{n!}.
\label{probability_equation}
\end{equation}

No galaxies are detected in our survey ($n=0$), so models which predict $\lambda_{\rm ex}$ detections can be rejected at the $(1-f_{n})$ level; i.e.\ a model which predicts 1 galaxy is rejected at the 63 per cent level ($\sim 1\,\sigma$ for a Normal distribution), and a model which predicts 3 galaxies are detected is rejected with 95\% confidence ($\sim 2\,\sigma$ for a Normal distribution).

As we did in Caruana et al.\ (2012), we compare our results with the work of Stark et al.\ (2010) at $4<z<6.5$ (the four data points with error bars in Figure \ref{fig:igm_constraint_combined}) whose spectroscopic sample covers a range in UV luminosities of -19.5 $>$ M$_{UV}$ $>$ -20.5, which is similar to the range of luminosities of our own sample.

As shown in Table \ref{table2_fors2}, the survey presented in this paper observed a probability-weighted sample of $N_{\rm eff}=8.46$ galaxies to a sensitivity threshold equivalent to 75\AA\, Equivalenth Width. If the fraction of galaxies with strong line emission at $z=7$ is consistent with the linear trend observed at lower redshifts (see Figure \ref{fig:igm_constraint_combined}), one would expect $X_{Ly\alpha}=0.25$. Thus, 2.12 galaxies in our sample would be expected to show Lyman-$\alpha$ emission at EW$\geq75$\AA. None are observed. Thus we rule out a scenario in which the Lyman-$\alpha$ fraction continues to evolve at the same rate at a confidence level of 88\%.

Since no confirmed spectroscopic redshifts for our $z>7$ sample are available, in drawing our conclusions we are assuming that our sample truly consists of $z>7$ galaxies; in other words, we are solely relying on the photometry to assume that the observed objects lie at $z>7$.  Given the very good photometric quality of WFC3 data and the fact that for our analysis weak candidates present in our sample were not considered, our observations are not expected to be compromised by a significant fraction of low-redshift contaminants.  Such contaminants are expected to be of two types: low-mass dwarf stars, especially T- and L-dwarfs (whose absorption features could resemble a spectral break), and low-redshift galaxies at $z\approx 1.5-1.9$.  In these cases, one might expect to detect some signature spectral line, such as [OII] for low-redshift galaxies and perhaps FeH for L-dwarfs.  These features would lie within our covered spectral range, but we do not observe any emission lines in any region of our spectra.  Furthermore, in the case of L-dwarfs, one might expect to observe a sharply rising continuum dominating the red end of the spectrum, which is not observed in our spectra.  This argues further against contamination by low-redshift sources.

Making this assumption, the data presented here imply that the fraction of star-forming $z=7$ galaxies with a Lyman-$\alpha$ Equivalent Width of 75\AA\, or greater is likely $<12\%$, rather than the $25\%$ predicted by extrapolations from lower redshift.  Indeed no strong evidence for Lyman-$\alpha$ emission is found in our spectroscopic sample of $z>7$ candidates.  

\subsection{Combining Results from All Spectrographs}

In total, our combined sample of targets across all spectrographs (i.e. the Gemini/GNIRS and VLT/XSHOOTER Observations discussed in Caruana et al.\ (2012) and our new FORS2 observations), consists of 27 $z$-drops and 3 $Y$-drops.  Ignoring any weak $z$-drop candidates in our FORS2 sample (7 objects), the revised total of targeted samples consists of 15 $z$-drops and 3 $Y$-drops.  Lyman-$\alpha$ emission was not detected in any of these objects.  In our analysis the fraction of Lyman-$\alpha$ emitters at $z>7$ is compared with that at lower-redshifts, based on the work of Stark et al.\ (2010), building on work by Stanway et al.\ (2007) at $z=6$ and Shapley et al.\ (2003) at $z=3$.  Accounting for known wavelength range restrictions and taking into consideration the completeness of our sample, effectively we observe 9.19 $z-$drops and 1.15 $Y$-drops with an Equivalent Width $<$ 75\AA.  

Extrapolating the fraction of Lyman-$\alpha$ emitters at lower redshifts to $z=7$, one expects $X_{Ly\alpha}\sim 0.25$ at $z=7$, i.e. the fraction of Lyman-$\alpha$ emitters is expected to be 0.25 at $z=7$.  Given our effective number of observed galaxies at $z=7$ (9.19) one would expect to observe $0.25\times9.19=2.3$ galaxies in the combined sample\footnote{We note that the analysis of Caruana et al.\ (2012) explored limits on 200\,km\,s$^{-1}$ lines, while the deeper FORS2 spectroscopy in this work has been examined for a broader 300\,km\,s$^{-1}$ line.} that is being considered, but none are detected (or at most one, if the tentative detection of ERSz-2225141173 which is marginally below our S/N cut, is considered to be real).  This implies that the hypothesised continued evolution in the emitter fraction can be rejected at a confidence level of $\sim 90\%$.

If the targeted $Y$-drops at $z=8.5$ are included in the analysis to derive a constraint at $z=7.8$ (the average redshift resulting from considering $z$-drops at $z\sim7$ and $Y$-drops at $z\sim8.5$) then one would expect $X_{Ly\alpha}\approx 0.3$ if the extrapolated evolution from lower redshifts of the Lyman-$\alpha$ fraction holds.  The total effective number of observed galaxies ($z$-drops and $Y$-drops combined) is 10.34 which means that across the whole sample one would expect to observe $0.3\times10.34=3.10$ galaxies.  Since none are observed, a model which predicts that the fraction of Lyman-$\alpha$ emitters at $z=7.8$ follows the trend seen at lower redshifts can be rejected at a confidence level of $\sim96\%$.

This decline in the detected frequency of Lyman-$\alpha$ emission in rest-UV selected star forming galaxies at $z=7$ is in agreement with that seen by other groups targetting both faint (eg. Fontana et al.\ 2010, Vanzella et al.\ 2011, Pentericci et al.\ 2011, Schenker et al.\ 2012) and relatively bright (Ono et al.\ 2012, using Subaru imaging) candidate objects. It is also supported by observations made at $z = 5.7$ and $z = 6.5$ by Ouchi et al.\ (2010) and Kashikawa et al.\ (2011) respectively, which report a decrease in mean Equivalent Width of Lyman-$\alpha$ at a fixed continuum luminosity with increasing redshift. The same study also reports little evolution in the rest-UV luminosity function for Lyman break galaxies whilst observing a decrease in the Lyman-$\alpha$ luminosity function. By contrast the work of Stark et al. (2010, extended in Stark et al.\ 2011) predicts that unless the neutral fraction rises between $z=6$ and $z=7$, Lyman-$\alpha$ should be readily detectable in spectroscopic campaigns at $z=7$. Such observations could imply an increase in $\chi_{HI}$, as discussed in the next section.

\subsection{Constraining $\chi_{HI}$, the Neutral Fraction of Hydrogen}
\label{sec:discussion}

\begin{figure*}
\begin{center}
   \resizebox{1.0\textwidth}{!}{\includegraphics{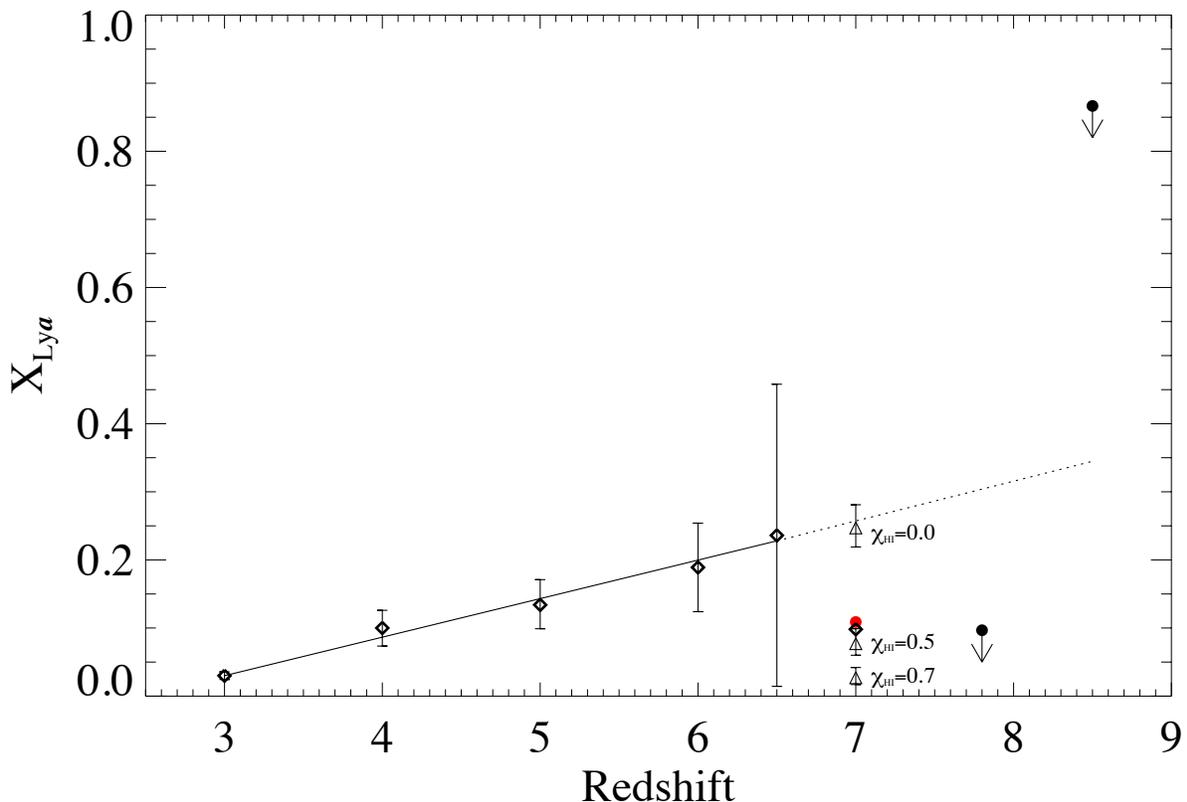}} \\
\end{center}
 \caption{Our upper limits on the fraction of high rest-frame equivalent width ($<75$\AA) Lyman-$\alpha$ emission at $z\ge7$ are shown for the $z$-drops (red datapoint at $z=7$) targeted with VLT/FORS2, Gemini/GNIRS and VLT/XSHOOTER and the $Y$-drops ($z=8.5$) targeted with VLT/XSHOOTER.  The upper limit at the average redshift of $z=7.8$ is for the total sample (including both $z$-drops and $Y$-drops).  The diamond symbol at $z=7$ is the actual fraction of Lyman-$\alpha$ emitters if the tentative detection of Lyman-$\alpha$ in ERSz-2225141173 is included in our sample, whereas the datapoint (filled, red circle) denotes the upper limit at this redshift.  The diamond symbols at $z<7$ are results obtained at lower redshift by Shapley et al.\ (2003) at $z=3$ and Stark et al.\ (2010) at $z=4-6.5$. For comparison, the low-redshift trend is extrapolated to higher redshifts (dotted line). Our upper limits are inconsistent with this extrapolation at a confidence level of 90\% at $z=7$ and 96\% at $z=7.8$.  Our results would suggest that the IGM neutral fraction, $\chi_{HI}\sim0.5$, which would be in agreement with other studies.} 
 \label{fig:igm_constraint_combined}
\end{figure*}

Numerous theoretical studies in the literature have investigated how an increase in the neutral fraction of Hydrogen in the Intergalactic Medium could impact the transmission of Lyman-$\alpha$ photons from galaxies (e.g.  Santos et al.\ 2004, Furlanetto et al.\ 2006, McQuinn et al.\ 2007, Mesinger et al.\ 2008).  Our full spectroscopic sample (combining the results of Caruana et al.\ 2012 and the current analysis) is compared to the simulations of McQuinn et al.\ (2007), which investigate the impact of patchy reionization on the Lyman-$\alpha$ line profile, the luminosity function and the clustering of Lyman-$\alpha$ emitters for various models of reionization.  In similar fashion to Stark et al.\ (2010), we derive a mapping between the Lyman-$\alpha$ fraction and $\chi_{HI}$, making the core assumption that the Lyman-$\alpha$ fraction evolves with redshift even if there is no change in the ionization state of the IGM, which assumption finds support in Bouwens et al.\ (2010), who suggest that obscuration by dust continues to evolve over $z\gtrsim6-8$.  This would imply that in the absence of any evolution in the ionization state of the IGM, the Lyman-$\alpha$ fraction would be expected to be even larger at $z\simeq 6.5-7$.  For this reason, like Stark et al.\ (2010), we extend the smooth evolution observed in the Lyman-$\alpha$ fraction between $z\sim4$ and $z\sim6$ to $z>6$ (see Fig. \ref{fig:igm_constraint_combined}).

One should note that the fraction of Lyman-$\alpha$ emitters over the $3.0 < z < 6.2$ redshift range does have a luminosity dependence, which is presented in Figure 13 of Stark et al.\ (2010).  However our comparison is ensured to be a fair one by considering only those galaxies in the sample which have a comparable absolute magnitude to the sample used in the analysis of Stark et al.\ (2010).  It should also be stressed that we are imposing the same rest-frame Equivalent Width threshold as that of Stark et al.\ (2010), namely, 75\AA.  Looking at samples of such objects at lower redshift, it is true that some sources might have lower Equivalent Widths, but these would be far below the sensitivity limits of this spectroscopy, so that subset of the population cannot be constrained. For this reason, the comparison which is being made here is between the high Equivalent Width emitters.

Figure \ref{fig:igm_constraint_combined} shows the location of various $\chi_{HI}$ constraints (triangle symbols) in order to  derive a comparison between our own results and these values to constrain the neutral fraction of Hydrogen in the Intergalactic Medium at $z=7$.  Our results at $z\sim7$ (marked with an arrow at $z=7$ in Fig. \ref{fig:igm_constraint_combined}) indicate evolution in $\chi_{HI}$ from lower redshifts, suggesting that $\chi_{HI}\sim0.5$ at $z\sim7$.  The combined limit from our $z$-drop observations and $y$-drop observations (arrow at $z=8.5$) is shown at the average redshift of $z=7.8$.

\section{Comparison with Other Studies and Alternative Interpretations}
\label{sec:comparison}

These results are in broad agreement with those obtained from other observational campaigns (Ota et al.\ 2008, Ota et al.\ 2010, Pentericci et al.\ 2011, Schenker et al.\ 2012, Ono et al. 2012).  A very recent paper (Finkelstein et al. 2013) reports only one instance of Lyman-$\alpha$ detection in a sample of 43 objects.  The object (z8\_GND\_5926 in their designation) is a very bright source, so if the observed line is indeed Lyman-$\alpha$, and especially if repeat observations yield the same result, this might indeed be suggesting that only objects that are luminous enough to clear an ionized zone around them in an increasingly opaque ISM can be observed.  On the subject of observations in support of an increase in $\chi_{HI}$, it is apt to also mention the quasar ULASJ112001.48$+$064124.3 (Mortlock et al.\ 2011), which lies at the low-redshift end of what is probed by our study.  The transmission cut-off of quasars at $z\sim6$, identified with the quasars' ionization fronts, led to estimates of the neutral fraction around quasars at $6.0 \lesssim z \lesssim 6.4$ is $\gtrsim 0.6$ (Wyithe et al.\ 2005).  From a measurement of the quasar's near-zone radius, Mortlock et al.\ (2011) find that the neutral fraction was a factor of about 15 higher at $z=7.1$ than what it was at $=6.2$.

In order to acquire an idea of how much support the drop in the Lyman-$\alpha$ fraction at $z>7$ has across all the published studies, we made a rough comparison with existing samples in the literature.  In order for such comparisons to be meaningful, it is important that the combined sample consists of objects having the same luminosity, so we have only selected targets that fall within the same luminosity range as those in our sample and that of Stark et al.\ (2010), namely $-20.5 < M_{UV} < -19.5$.  This enabled us to add a further 15 unique (i.e. not multiply observed) objects to our own, and these are listed in Table \ref{table:compwithlit}.  Comparing observations across different observing setups can be difficult, so we have carried out our estimate for two representative cases, each assuming a different completeness fraction.  In the first case, we assume a somewhat conservative 60\% completeness fraction, which is similar to our own.  This increases the total sample size at $z=7.8$ to 19.34, out of which only 2 objects are detected, namely NTTDF-474 by Pentericci et al. (2011) and ERS8496 by Schenker et al. (2011).  (Note that we are not including here HUDF09\_1596 by Schenker et al. 2011, which they consider to be a marginal detection.)  In this case, the rising trend of the Lyman-$\alpha$ fraction observed at lower redshifts can be rejected only at the 95\% confidence level.
On the opposite side of the scale, if we assume (unrealistically) a 100\% completeness fraction for all the other samples in the literature, then the total sample size would consist of 25.34 objects and the confidence level at which we can reject the trend observed at lower redshifts would rise to 98.6\%, which still remains below the 3$\sigma$ level.  This simple exercise indicates that whilst significant effort has gone into spectroscopic observations at $z\sim7$ by various groups (see also recent work by Tilvi et al. 2014, Faisst et al. 2014, Pentericci et al. 2014 and Schenker et al. 2014) and a lot of progress has been made from an imaging point of view at $z\sim8$, acquiring more spectroscopic data to assess the distribution of Lyman-$\alpha$ emission and absorption at these redshifts is a worthwhile endeavour.  These multiple observational efforts suggest a decline in the fraction of Lyman-$\alpha$ emitters at $z>7$, but to confirm this at a high level of statistical significance, larger samples are required.  

\begin{table*}
\centering
\begin{tabular}{ | c | c | c |}
\hline
Observed Target & Absolute Magnitude\\
\hline
\hline
Study by Fontana et al. (2010) & \\
\hline
G2\_4034 & -20.50 \\
\hline
Study by Schenker et al. (2011) & \\
\hline
ERS5847 & -20.22 \\
ERS7412 & -19.89 \\
ERS8119 & -19.79 \\
ERS8290 & -19.73 \\
ERS8496* & -19.51 \\
ERS10270 & -19.54 \\
A1703\_zD1 & -20.39 \\
HUDF09\_1584 & -20.27 \\
HUDF09\_1596 & -20.12 \\
\hline
Study by Pentericci et al. (2011) & \\
\hline
NTTDF-474* & -20.35 \\
NTTDF-1632 & -20.45 \\
NTTDF-2916 & -20.25 \\
BDF4-5583 & -20.24 \\
BDF4-5665 & -20.25 \\
\hline
\hline
\end{tabular} 
\caption{The sample of unique (i.e. not multiply observed) $z$-drops from the literature with luminosities lying within the range of our own objects which we added to our own sample.  The two objects marked with an asterisk have detectable Lyman-$\alpha$ emission.}
\label{table:compwithlit} 
\end{table*}

A study at $z=6.5$ by Cowie et al.\ (2011) highlights the importance of further investigation.  Cowie et al.\ (2011) focused on 7 narrow-band selected galaxies at $z=6.5$ (which were selected from the Lyman-$\alpha$ spectroscopic atlas of Hu et al.\ 2010); comparing these with continuum-selected samples at the same redshift, they report that the fraction of Lyman-$\alpha$ emitters is similar to that at lower redshifts, suggesting that they find no sign of the effects associated with reionization.  So whilst, as has seen above, numerous studies do seem to be in agreement with our own findings that there is a drop in the Lyman-$\alpha$ fraction, it is very important to increase the sample size in order to determine whether results such as Cowie et al.'s (2011) might be suffering from uncertainties arising from cosmic variance in the continuum luminosity functions, or from cosmic variance in the case of patchy reionization.  Also, we note that since the study of Cowie et al.\ considers a slightly lower redshift ($z=6.5$ as opposed to $z>7$), this discrepancy might simply be indicating that most evolution takes place at $z>6.5$. Such continued study would also prove useful in investigating other interesting questions, such as the homogeneity of the reionization process.

Dunkley et al.\ (2009) use polarization measurements from WMAP data to derive constraints for reionization, claiming that instantaneous reionization during late epochs below $z=8.2$ (6.7) can be rejected at the $2\sigma$ ($3\sigma$) level, arguing instead for an extended reionization process taking place between $z\sim6-11$.  Extended reionization is also supported by the newer Planck results (Planck collaboration and Ade et al. 2013).  The current problem is that physical models cannot easily reproduce such an extended era of reionization because of the rapid recombination of Hydrogen (e.g. Cen et al.\ 2003).  Studies such as the one presented here can be used to shed some light upon the question of when and over how long a period reionization took place.  Indeed, if the lack of observed Lyman-$\alpha$ emission at $z=7$ is interpreted as signifying an increase in the neutral fraction from lower redshifts, suggesting a significant departure from $\chi_{HI}=0$, then it would be telling us that reionization was definitely not complete by $z=7$, constraining the redshift range over which reionization could have occurred.  Further spectroscopic studies on even larger samples of high-redshift galaxies hold the promise of tackling this puzzling question with greater clarity and possibly solving this cosmological problem.

Finally, it should be noted that there are various theoretical discussions in the literature about the interpretation of the observed drop in Lyman-$\alpha$ emitters at high redshift, and their use as a tool to constrain the neutral fraction.  Bolton \& Haehnelt (2013) suggest an alternative viewpoint as to whether this drop in the Lyman-$\alpha$ fraction should indeed be interpreted as signifying a change in the neutral fraction from that at lower redshifts. Employing a hydrodynamical simulation, they show that towards the tail-end of reionization, in average parts of the universe the opacity of the IGM red-ward of rest-frame Lyman-$\alpha$ can increase significantly due to a higher number of optically thick absorption systems; a more modest rise in the volume-averaged neutral fraction, even as low as 10\%, would be sufficient to reduce the transmission redward of rest-frame Lyman-$\alpha$.  Dijkstra et al. (2014) argue for the importance of considering an evolving ionizing photon escape fraction, and Mesinger et al. (2014) predict that IGM attenuation alone would not result in a Lyman-$\alpha$ fraction drop larger than a factor of about 2, indicating that any further evolution would require evolution in the properties of the galaxies themselves.  However, it should be stressed that the size of the current observed sample is not sufficiently large to address the validity of these constraints.

\section{Conclusions}
\label{sec:conclusions}

We have presented FORS2 spectroscopy of 22 $z$-band dropout galaxies and 16 $i$-band dropout galaxies.  We detect Lyman-$\alpha$ emission in 4 $i$-drops and a tentative emission line in one $z$-drop, but do not find any evidence for Lyman-$\alpha$ in the rest of the sample.  Our observations of HUDF.z.4444, a priorly observed object by Fontana et al. (2010, G2\_1408 in their catalogue), which was claimed to exhibit tentative Lyman-$\alpha$ emission at a wavelength of 9691.5\AA\, (corresponding to $z=6.97$), rule out this emission claim at a level $>3\sigma$ ($>$ 99\% probability).  Our observed Lyman alpha fraction is $<12\%$ at $z=7$, lower than that predicted by a factor of 2.  Drawing comparisons with theoretical studies, we constrained the neutral fraction of Hydrogen, $\chi_{HI}$, at z$\sim$7 to be $\sim 0.5$. 

We are living in very interesting times for studies of the high-redshift universe, both from an observational and theoretical perspective.  Currently, the sample of candidate galaxies at $z>7$ is fairly large, however, the number of spectroscopically confirmed objects remains very small.  It would be a very worthwhile endeavour to extend spectroscopy to larger samples, in order to investigate whether these conclusions still hold.

No doubt, theory and observation will continue to inform each other in the coming years.  From a practical point of view, studies such as the one presented in this paper show that such observations are notoriously difficult with current instrumentation, because it is challenging to obtain large samples at $z>6.5$, especially because of the deteriorating effect of atmospheric emission lines, which hamper such observations and limit the level of completeness.

New instruments such as KMOS at the VLT will be well suited for IR follow-up surveys of such high-redshift objects, and ULTRA-VISTA will probe the bright end of the luminosity function, making it a perfectly poised instrument for selecting candidate galaxies suitable for follow-up spectroscopy.  Follow up spectroscopy with MOSFIRE at Keck has the potential to provide the first evidence of Lyman-$\alpha$ emission at $z\sim8$, although first attempts have not yielded any detections yet (Treu et al.\ 2013), while the VLT IFU spectrograph MUSE will allow us to look for Lyman-$\alpha$ emission over a wide redshift range (up to $z\sim6.7$) and would be particularly useful in improving current constraints on the faint end of the Lyman-$\alpha$ emitting galaxy luminosity function and in improving measurements of the Lyman-$\alpha$ fraction at $z=6-6.5$.

In the absence of spectroscopic confirmation, current estimates of the neutral fraction of Hydrogen are based on an estimated redshift range.  In the long term, the James Webb Space Telescope (JWST) will be a workhorse instrument for both identifying ultra high-redshift galaxies and also confirming them spectroscopically with the state-of-the-art multi-object-spectrograph NIRSpec, which will allow us to obtain spectroscopy on more than 100 sources simultaneously.  If Lyman-$\alpha$ is indeed not emerging at these high redshifts, then spectroscopic confirmation of high-$z$ candidates could be achieved using the [OII] line or other lines, including rest-UV absorption in the Interstellar Medium and stellar atmospheres.  JWST will be able to confirm the redshift distribution of these sources, and finally rule out the presence of faint contaminants, hence allowing more precise estimates of the neutral fraction to be made.

\subsection*{Acknowledgements}

Based on observations made with the NASA/ESA {\em Hubble} Space Telescope associated with programme \#GO-11563,
obtained from the Data Archive at the Space Telescope Science Institute, which is operated by the Association
of Universities for Research in Astronomy, Inc., under NASA contract
NAS 5-26555. 
We thank the anonymous referee for their useful comments which improved this paper.
We gratefully acknowledge Lance Miller, Malcolm Bremer and Mark Lacy for useful discussion and comments.
For this work, JC and SL were supported by the Marie Curie Initial Training Network ELIXIR of the European Commission under
contract PITN-GA-2008-214227.  SL acknowledges support from the Science and Technology Foundation (FCT, Portugal) through 
the fellowship SFRH/BPD/89554/2012 and the research grant PEst-OE/FIS/UI2751/2011.
AB and SW acknowledge financial support from an STFC Standard Grant.

\end{document}